\documentclass[aps,preprint,showpacs,superscriptaddress,groupedaddress,floatfix]{revtex4}  
\usepackage{graphicx}  
\usepackage{float}
\usepackage{dcolumn}   
\usepackage{bm}        
\usepackage{amssymb}   
\usepackage{amsmath}
\usepackage{isomath}
\usepackage[usenames, dvipsnames]{color}
\usepackage{color,soul}
\soulregister\cite7
\soulregister\ref7
\soulregister\pageref7
\usepackage{xr}
\usepackage{chngcntr} 
\usepackage{secdot} 
\usepackage{soul} 

\def\hk#1{{\color{cyan} #1}}

\begin{document}

\title{Bloch oscillation of elastic waves in the graded lattice of 3D-printed hollow elliptical cylinders}
\author{H. Kim}
\affiliation{Aeronautics and Astronautics, University of Washington, Seattle, WA, 98195-2400, USA}
\author{X. Shi}
\affiliation{Aeronautics and Astronautics, University of Washington, Seattle, WA, 98195-2400, USA}
\author{E. Kim}
\affiliation{Division of Mechanical System Engineering \& Automotive Hi-Technology Research Center, Chonbuk National University,567 Baekje-daero, Deokjin-gu, Jeonju-si, Jeollabuk-do 54896, Republic of Korea}
\author{J. Yang}
\affiliation{Aeronautics and Astronautics, University of Washington, Seattle, WA, 98195-2400, USA}

\date{\today}

\begin{abstract}
We study the Bloch oscillation of elastic waves in a chain composed of hollow elliptical cylinders (HECs). These HECs are 3D-printed in different wall thicknesses and are arranged to form a graded chain. 
We find that the frequency band structure of this lattice can be manipulated in a way to create a narrow strip of transmission range sandwiched between slanted stop bands. This enables the trapping of elastic waves at a specific location of the chain, which depends on the input frequency of the propagating elastic waves. 
This elastic Bloch oscillation in a tailorable 3D-printed system enables the control of energy localization in solids, potentially leading to engineering applications for vibration filtering, energy harvesting, and structural health monitoring.

\end{abstract}

\maketitle


\section{Introduction}
Bloch oscillation describes interesting quantum mechanics of electrons. When an external electric field is applied at a constant power, electrons in a periodic potential show spatially localized oscillations \cite{Bloch1929,Zener1934}. The Bloch oscillation is caused due to the equidistant energy band known as the Wannier-Stark ladder \cite{Wannier1960}. In essence, it is the frequency-domain counterpart of the Bloch oscillation.
Despite the prediction of the Bloch oscillation in the early twentieth century, it was not until the end of the century that researchers finally observed it. Thanks to the advent of semiconductor superlattices \cite{Esaki1970}, the Wannier-Stark ladders \cite{Mendez1988,Bleuse1988} and the Bloch oscillations \cite{Kuzmin1991,Feldmann1992,Leo1992,Waschke1993,Dekorsy1994,Martini1996,Loser2000}  were experimentally demonstrated. 

Later on, researchers in other field of physics started to show significant interest in the Bloch oscillation. Optical Bloch oscillations have been reported over the last couple of decades \cite{Pertsch1999,Morandotti1999,Sapienza2003,Agarwal2004,Lousse2005,Ghulinyan2005,Joushaghani2009}. The acoustic counterpart has also been explored \cite{Kimura2005,Sanchis2007,Kimura2010,He2007,Lima2010,Karabutov2013,Lazcano2014}. 
However, there is a very limited number of published works on the Bloch oscillations of mechanical waves. Gutierrez \emph{et al} \cite{Gutierrez2006} studied the Bloch oscillations of torsional waves traveling along the arrays of elastic rods in geometric gradient, which mimicked the effect of the electric field on electrons. Arreola-Lucas \emph{et al} \cite{Arreola-Lucas2017} also observed the Bloch oscillation of torsional waves passing through a metallic beam with notches which has a gradient in the cavity thickness. In both cases, researchers manufactured their systems by machining metal, which is not favorable for exploring their systems thoroughly by changing design variables. Recently, Shi \emph{et al} \cite{Shi2018} came up with a tunable system which realizes the Bloch oscillations of stress waves. They used a chain of solid cylinders with the gradient in their contact angles. This eventually results in the gradient of contact stiffness which creates the Wannier-Stark ladder.

To achieve tunability of the system, 3D-printing can provide a powerful solution. 3D-printing gives users a variety of options in terms of the material, precision, and scale. We find many fields of study or industry incorporating the 3D-printing technique, and the elastic wave community is not an exception. Examples include shock absorbing structures \cite{Tsouknidas2016,Bates2016,Chen2018}, jumping robots using soft materials \cite{Bartlett2015}, compliant structures for carrying solitons \cite{Deng2017},  energy trapping \cite{Shan2015} and acoustic filters devices \cite{Li2016}.

In this Letter, we numerically and experimentally verify the Bloch oscillation in the 3D-printed chain of hollow elliptical cylinders (HECs). We impose a linear gradient to the thickness of the HECs in the chain. 3D-printing provides an exceptional tunability in the design and fabrication of the HECs. With the power of 3D-printing, we can manufacture the HECs at desired thicknesses with high precision. The controlled thickness variation results in the gradient in the contact stiffness \cite{HKim2018Graded}, creating slanted frequency band structures. This can form a counterpart of the
Wannier-Stark ladders for elastic waves. 
As a result, we observe the Bloch oscillations of the waves in the mechanical test bed. We show that the location of the Bloch oscillations depends on the excitation frequency due to the Wannier-Stark ladders. This enables the control of energy localization in solids for potential engineering applications.

The rest of the manuscript is arranged in the following order. We first explain experimental methods and simulation details in Section \ref{Sec:method_lin}. Then in Section \ref{Sec:Result_lin}, we discuss the frequency bands of the graded HEC chain and how it affects the frequency response of the chain. Finally, we show the dynamic response of the graded HEC chain to various excitation frequencies and demonstrate the Bloch oscillation in the chain. We finish the manuscript with conclusions. 

\section{Methods}\label{Sec:method_lin}
\begin{figure}[t]
\centering
\includegraphics[width=.75\linewidth]{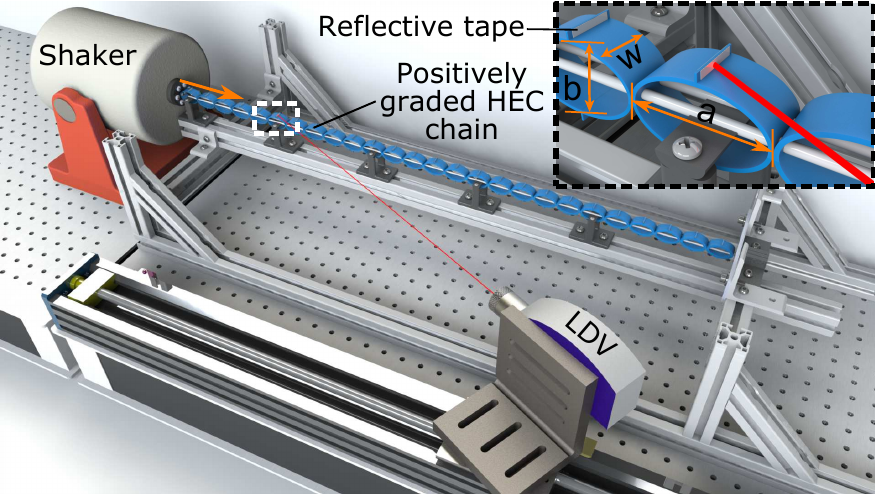}
\caption{A schematic diagram of the experimental setup for the linear perturbation system. One end of the HEC chain is mounted to the shaker head to excite the system longitudinally (orange arrow). The graded HEC chain is securely positioned in the frame through 3D printed jigs. We measure the velocity of each HEC by focusing the laser beam on the reflective tape (the inset on the top right corner, a zoomed-in view of the dashed box in the main image). The dimensions of the HEC are $a$ = 30 mm, $b$ = 18 mm, and $w$ = 12 mm as shown in the inset.} 
\label{fig:exp_setup_shaker}
\end{figure}

We 3D-print HECs with polylactic acid (PLA) material (Black, Ultimaker). The wall thickness of HECs varies linearly from 0.4 mm to 3 mm in 26 steps. Then we assemble these 26 HECs into a chain, as shown in Fig.~\ref{fig:exp_setup_shaker}. HECs are aligned through two stainless steel shafts to confine their movement to sliding in the major axis. At the contact point, the HECs are bonded using super glue (Loctite 431) to secure their contact. The first HEC is tightly bolted into the shaker head where the input signal is applied. We apply a precompression of 10 mm to the entire chain and excite the chain at a low amplitude to ensure that the elastic wave is in the linear regime. 

We send a frequency sweep signal or a Gaussian pulse to the shaker through a function generator (33220A, Agilent). When the chain is excited, the laser Doppler vibrometer (LDV; Polytec OFV-534) measures the velocity of a single HEC (inset of Fig.~\ref{fig:exp_setup_shaker}). The velocity data is acquired by the oscilloscope (DSO-X 3024A, Agilent) and downloaded to a local computer for post-processing. We repeat the experiment five times for statistical treatments. 

We use commercial finite element analysis (FEA) software (ABAQUS) to perform numerical analysis. We select the 2D quadratic Timoshenko beam (B22 element) to model the HECs.
We simulate only the upper half of the HECs, assuming symmetry \hk{about} the axial direction. We apply the Neo-Hookean model for implementing the hyperelastic property of the PLA material. We find the material constant of the Neo-Hookean model by fitting to the experimental data ($\mu=800\times 10^6$ N/m$^2$).  
We ignore the material damping in the simulations.

\section{Results and Discussion}\label{Sec:Result_lin}
\begin{figure}[t]
\centering
\includegraphics[width=1\linewidth]{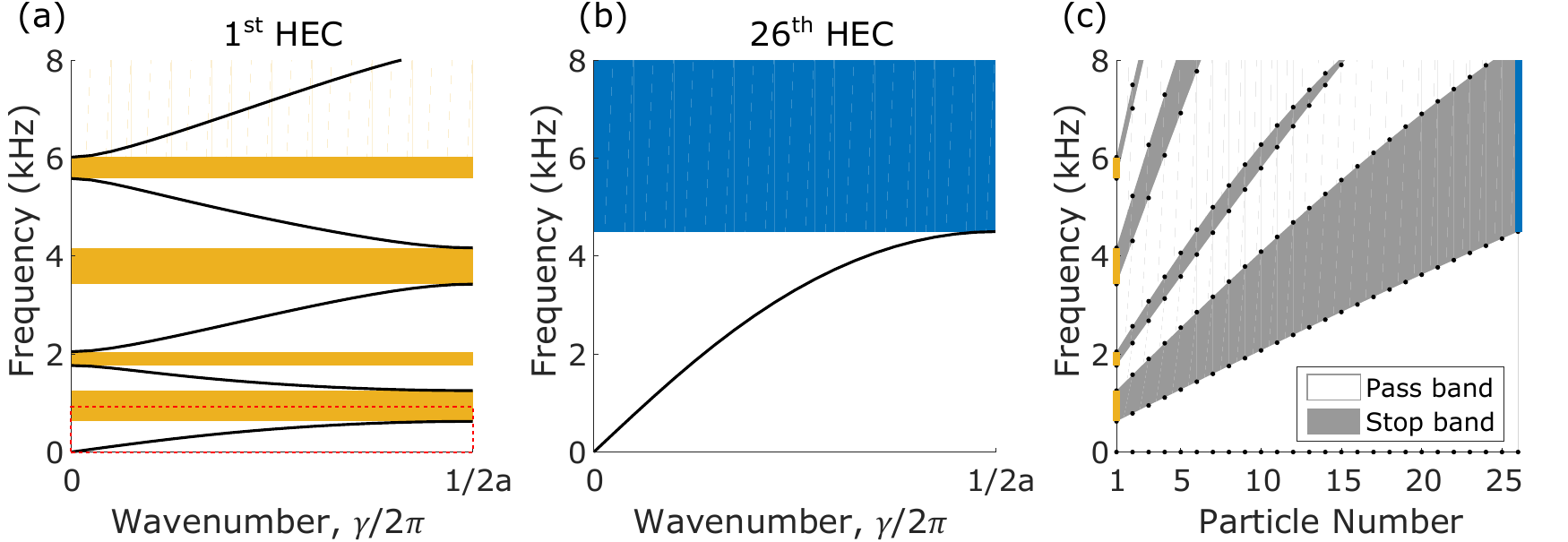}
\caption{Dispersion curves for (a) the fist HEC (thickness of 0.4 mm) and (b) the last HEC (thickness of 3 mm) in the graded chain obtained using FEA. The shaded areas represent stop bands where wave does not propagate, whereas the white area is the pass band where wave transmits. The red box in (a) shows similarity with (b). (c) Band structure of the graded HEC chain. The gray areas indicate the stop band. The yellow and the blue bars come from (a) and (b). 
}\label{fig:DispCurve}
\end{figure}

We first investigate the band structure of the graded HEC chain. By assuming that the unit cell analysis can approximately estimate the \emph{local} frequency band, we assemble the band edge frequencies to build the band structure of the graded chain. 
Figure~\ref{fig:DispCurve} shows how the band structure of the graded HEC chain is constructed. 
The first HEC exhibits multiple narrow pass and stop bands (Fig.~\ref{fig:DispCurve}(a)). This is because the thin 3D-printed cylinder has multiple eigen frequencies and corresponding eigen modes within the frequency range of interest. The frequency bands move to a higher range as the thickness of the HEC increases due to increase of the bending stiffness associated with the eigen frequencies. The last HEC only has a single pass/stop band within the frequency range of interest. It is notable that the pass/stop band of the last HEC in Fig.~\ref{fig:DispCurve}(b) corresponds to the the red dashed box in Fig.~\ref{fig:DispCurve}(a).
In essence, the dispersion relation of the HECs in the graded chain is nearly a scale-up of that of the first HEC. We only linearly increase the thickness of the HECs and keep everything else the same. Therefore, it is fair for us to expect approximately proportional increase in the natural frequencies. 
 As a result, if we calculate the dispersion relations of all of the individual HECs and assemble them in order, it creates slanted lines of pass bands and stop bands, as shown in Fig.~\ref{fig:DispCurve}(c). This resembles the tilted energy band for the electron cases. 

\begin{figure}[t]
\centering
\includegraphics[width=1\linewidth]{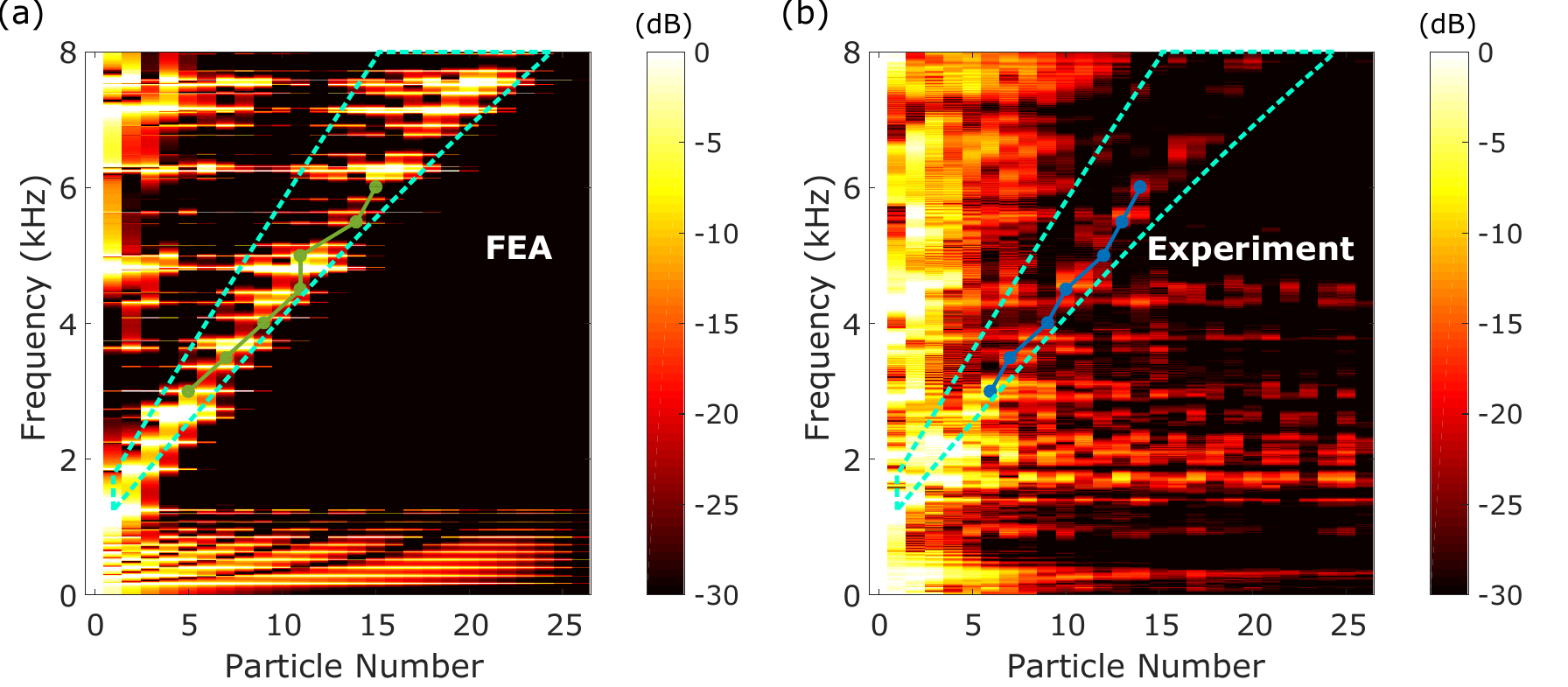}
\caption{Transmission of the velocity signal (a) from the steady-state response using FEA simulation and (b) under the chirp input from the experiment. Brighter color means higher intensity of the signal. The the FEA and the experimental data from Fig.~\ref{fig:Vmap}(d) are superimposed on top of (a) and (b), respectively. The second lowest pass band in Fig.~\ref{fig:DispCurve}(c) is superimposed on (a) and (b) in the dashed cyan box.
}\label{fig:sweep}
\end{figure}

Given the interesting slanted band structure, we are curious about the response of the
graded HEC chain to single-frequency excitations. 
To find that, we examine the frequency response of the chain by conducting FEA. We perform steady-state dynamic analysis and calculate the transmission. Figure~\ref{fig:sweep}(a) shows the transmission level in dB from the FEA result. We observe a transmission band with a positive slope develops from around the origin. We verify the FEA result by conducting experiment. We apply a chirp signal ranging linearly from 0.3 kHz to 8 kHz to the left boundary of the chain and measure the velocity for each HEC. We then perform fast Fourier transform (FFT) on the output data to analyze the frequency components. We observe similar behavior in the experimental result compared to the FEA result, as shown in Fig.~\ref{fig:sweep}(b). 
 While the highlighted branch continues growing up in the high frequency range, its intensity diminishes. It could be mainly due to the positive gradient of the HEC thickness in the chain. The particle velocity has to decrease as the HEC mass increases to meet the conservation of momentum. Another reason could be the large damping in the high frequency domain. 

Let us relate the frequency response (Fig.~\ref{fig:sweep}(b)) to the band structure (Fig.~\ref{fig:DispCurve}(c)). 
 We observe a near total transmission for the low frequency range (Fig.~\ref{fig:sweep}(b)) due to the complete pass band throughout the chain (Fig.~\ref{fig:DispCurve}(c)). The second pass band is positively sloped and sandwiched between the stop bands with the lower stop band being very thick. The wave can pass through the initial narrow stop bands because the evanescent wave width is possibly wider than the band width. Once the wave reaches the thick stop band, it cannot propagate further. This is what results in the narrow and sloped transmission band in the frequency response (the dashed cyan boxes in Fig~\ref{fig:sweep}).
What this interesting frequency response means is that we have frequency-dependent localization in the graded HEC chain. The narrow transmission band sprouting positively in the frequency-space map causes the wave captured in farther locations at higher frequencies. 

\begin{figure}[t]
\centering
 \includegraphics[width=1\linewidth]{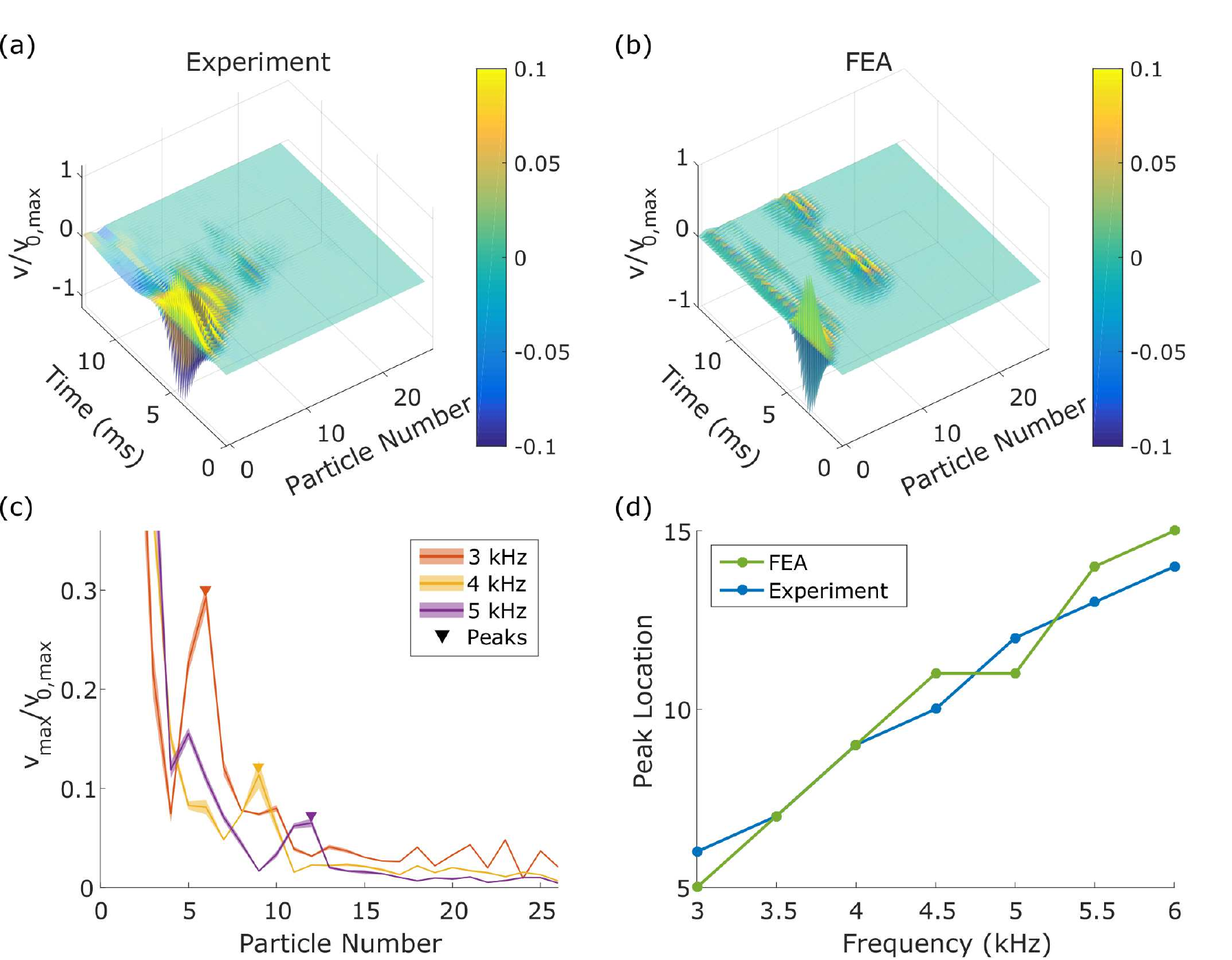}
 \caption{Velocity response to the input frequency of 5 kHz mapped in time and space domain, from (a) the experiment and (b) the FEA simulation. The velocity is normalized by the maximum input velocity ($v_{0,max}$). (c) The maximum velocity at each particle number is plotted at 3, 4, and 5 kHz input frequencies in orange, yellow, and purple solid line, respectively. The local peaks are marked with inverted triangles of respective color. The shaded areas represent standard deviations from the experiment. (d) Localization at input frequencies from 3 kHz to 6 kHz with step size of 0.5 kHz, from the experiment (blue line) and the FEA (green line).}
 \label{fig:Vmap}
\end{figure}

To verify that, we analyze transient response of the graded HEC chain. We apply a Gaussian pulse with a central frequency of 5 kHz to the left end and measure the response of each HEC. The result is plotted in Fig.~\ref{fig:Vmap}(a). The input signal quickly disappears but reappears around the 12th HEC. This signal is trapped and oscillates in this location for around 2 ms and starts to die out. This is a very narrow Bloch oscillation. 
We observe similar behavior from the FEA in Fig.~\ref{fig:Vmap}(b). The localization of the wave in the middle of the chain keeps its height for a longer time in the simulation. Had we not had friction in the experiment, we would ideally achieve this clear trapping. Overall, we confirm qualitative agreement of the localization effect in the graded HEC chain between the experiment and the FEA.

Next, we investigate the effect of the input frequency on the localization behavior. To achieve that, we conduct the same procedure at different input frequencies and plot the maximum velocity profile, as shown in Fig.~\ref{fig:Vmap}(c). It is evident that the localization peak is positioned farther at higher frequency. 
The peak amplitude decreases as the input frequency increases. We have pointed out earlier (Fig.~\ref{fig:sweep}) the reason includes the gradient in mass and the material damping.
Figure~\ref{fig:Vmap}(d) shows the localization point with respect to a wide range of input frequencies. Both the experimental data and the FEA result show positive relationship between the input frequencies and the peak locations. They also draw almost linear curves. This aligns very well with the transmission band we found earlier, as shown in the blue and the green curve in Fig.~\ref{fig:sweep}(a) and (b), respectively. In the Appendix A, we explore the tunability of the HEC system to create nonlinear evolution of band structures.

\section{Conclusion}
In summary, we investigated the elastic Wannier-Stark ladders and the Bloch oscillations in a 3D-printed, graded HEC chain. We experimentally and numerically demonstrated that the location of the Bloch oscillation depends on the input frequency in our system. Within the frequency range of interest, we find that the higher the input frequency is, the farther the localization happens. This is due to the slanted band structure, or the Wannier-Stark ladders, of the graded HEC chain. 

This study showcases that the design boundary for the band structure (i.e., bandgap) engineering can be broadened by the soft lattices made of 3D-printed architectures. This can enable an enhanced degree of freedom in controlling stress waves in solids, thereby realizing a plethora of novel wave dynamics. While this study focused on the Bloch oscillations in the graded lattice, the same system in the opposite gradient can provide another interesting phenomenon, so-called the rainbow (or boomerang) effect (Appendix B). 

The findings from this study can be integrated into various engineering applications, such as energy harvestor, structural health monitoring, and nondestructive evaluation systems, which require energy trapping. We also see the possibility of shedding light on the design of artificial cochlea, as it captures the wave of different frequency at different locations. 

\section*{Acknowledgments}  
The authors acknowledge fruitful discussions with Panayotis Kevrekidis from the University of Massachusetts, Amherst. J.Y. thanks the support of the National Science Foundation under Grant No. CAREER-1553202. E.K acknowledges the support from the National Research Foundation of Korea (NRF) grant funded by the Korea government (MSIP) under Grant No.2017R1C1B5018136.


\clearpage
\appendix
\raggedbottom\sloppy
\counterwithin{figure}{section}

\section{Tunability of the 3D-printed graded HEC chain}\label{ApdxB}
The 3D-printed HEC chains are a highly tunable system, enabling the manipulation of the cell geometry to control their dynamic response. In this section, we explore two possible examples to change the dynamic response of the chain: a) the cell geometry and b) the gradient. The band structure changes depending on the aspect ratio of the unit cell, which ultimately will change the localization position. We can also give a gradient not only in a linear manner but also in nonlinear patterns. This way, we can tune the the frequency-localization relationship from linear to nonlinear.

\begin{figure}[ht]
\centering
 \includegraphics[width=1\linewidth]{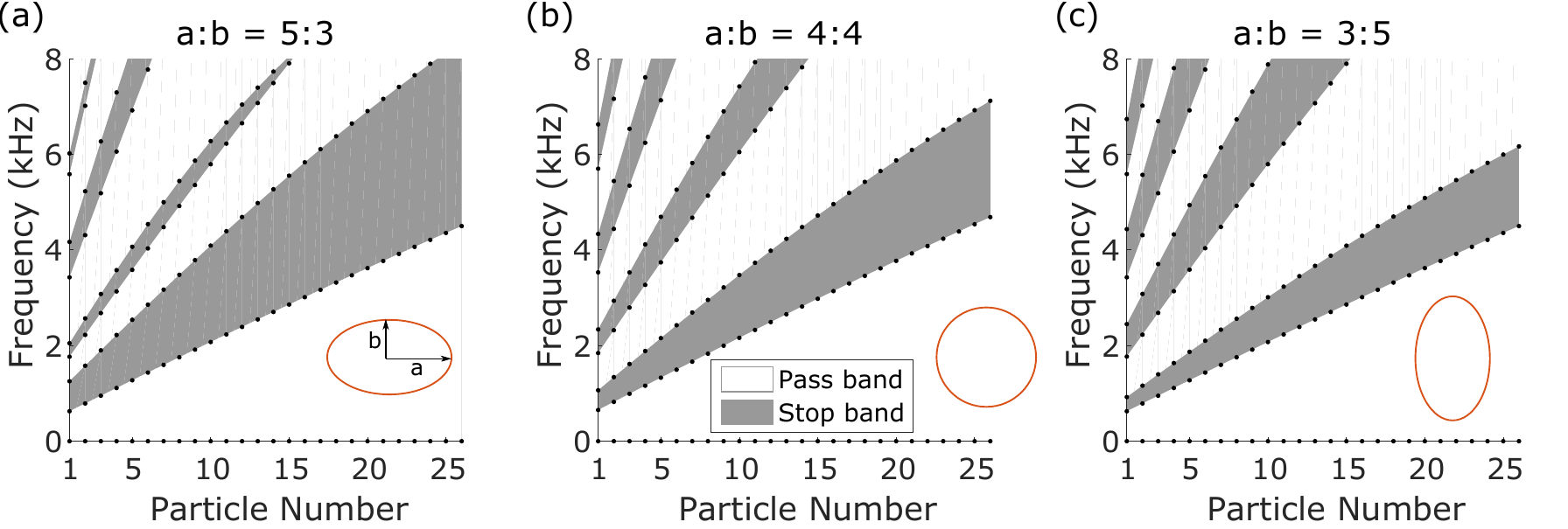}
 \caption{The band structure of the graded HEC chain whose aspect ratio, a:b (longitudinal to transverse), is (a) 5:3, (b) 4:4, and (c) 3:5. 
 The insets are the HEC configurations.}
 \label{fig:aratio}
\end{figure}

First, we study the effect of the aspect ratio of the HEC on the dispersion relation of the graded HEC chain. In the main manuscript, we only show the result for the aspect ratio of 5:3 which is repeated in Fig.~\ref{fig:aratio}(a). The aspect ratio is defined as the longitudinal to the transverse direction length. If we decrease the aspect ratio, i.e., change the ellipse to a circle (Fig.~\ref{fig:aratio}(b)) and to the other side of the ellipse (Fig.~\ref{fig:aratio}(c)), we see the first thick stop band shrinks and the second thin stop band expands. Consequently, the slanted pass band widens. As a result, we expect to observe clear Bloch oscillations as the wave is localized within a wide length of the chain.

\begin{figure}[t]
\centering
 \includegraphics[width=1\linewidth]{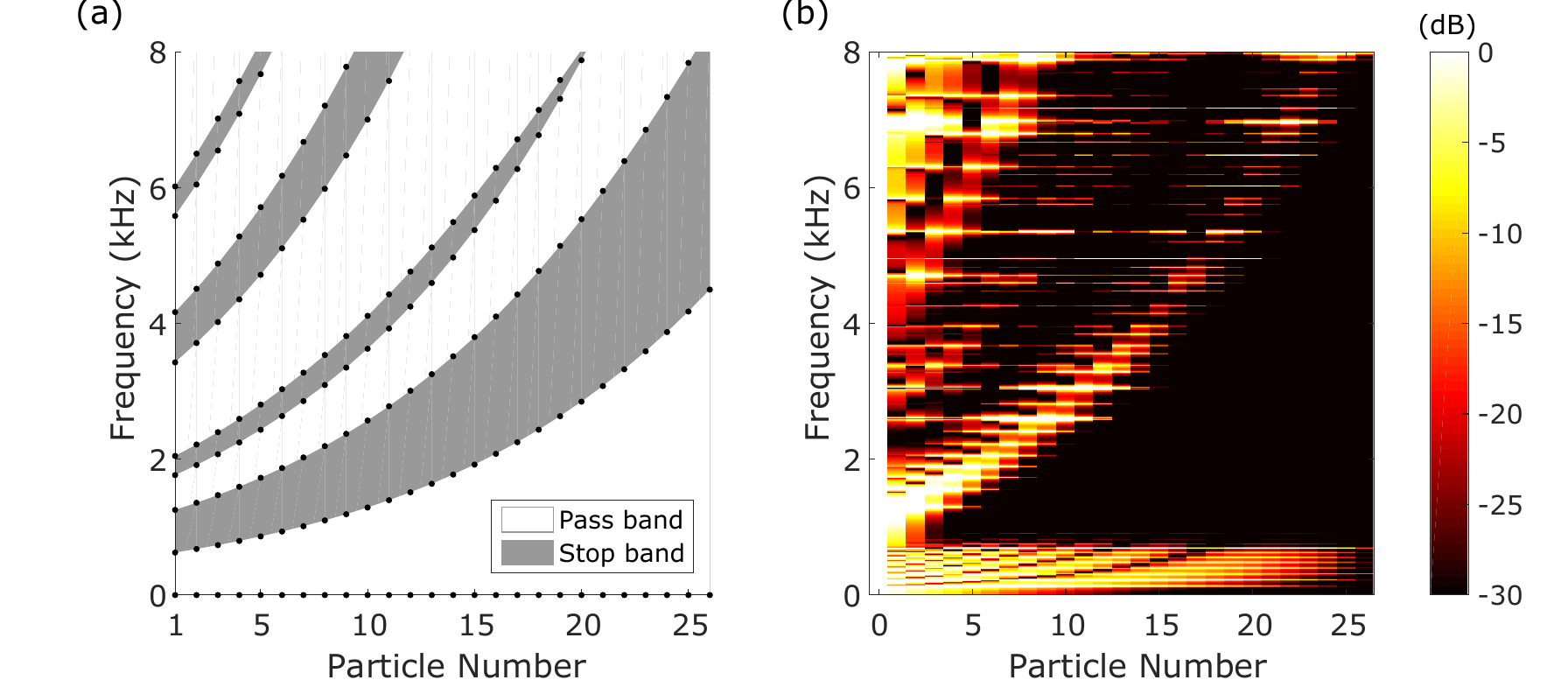}
 \caption{(a) Band structure of the HEC chain with exponential thickness gradient. Wave can only travel in the pass band. (b) Frequency response of the same chain using steady-state analysis. }
 \label{fig:exponential}
\end{figure}

Next, we investigate the effect of the gradient on the dynamics of the graded HEC chain. We vary the thickness exponentially from 0.4 mm to 3 mm and calculate the dispersion relation for each HEC. Then we construct the band structure by assembling the individual dispersion relations, as shown in Fig.~\ref{fig:exponential}(a). It is evident that the pass and the stop bands are curved rather than straight (Fig.~\ref{fig:sweep}(c)) in spatial domain. As a result, we observe a curved transmission branch in the frequency response in Fig.~\ref{fig:exponential}(b). This shows the capability of the graded HEC system to change the localization point by tuning the gradient parameters.

\section{Negative thickness gradient}\label{ApdxA}
\begin{figure}[ht]
\centering
 \includegraphics[width=1\linewidth]{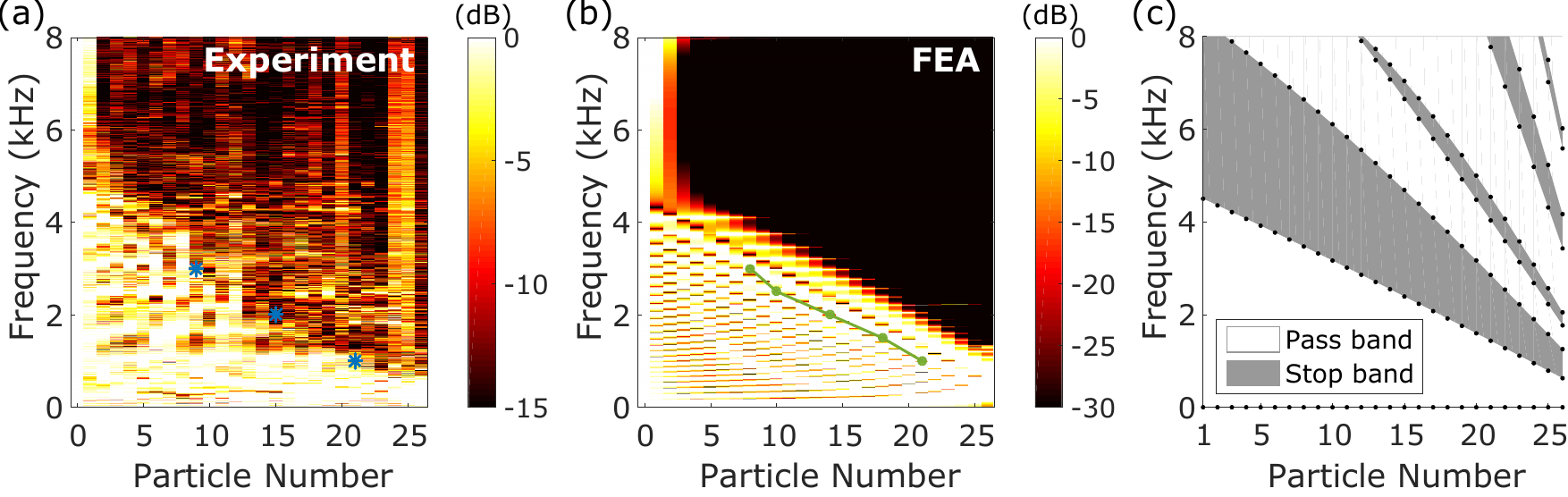}
 \caption{Transmission of the velocity signal of the HEC chain with negative gradient (a) under the chirp input from the experiment and (b) from the steady-state response using FEA simulation. The blue markers and the green line are superimposed from Fig.~\ref{fig:Vmap_thinning}(d). (c) Band structure of the negatively-graded HEC chain. The gray area indicate the stop band where wave does not propagate, whereas the white area is the pass band where wave transmits.}
 \label{fig:sweep_thinning}
\end{figure}

In the main manuscript, we only discuss the HEC chain with positive thickness gradient. We can easily investigate linear dynamics of the negative gradient HEC chain by simply flipping the chain in the opposite direction. In this section, we discuss how the dynamics changes in the negative thickness gradient.
We measure the frequency response of the chain and compare it with the FEA result and the band structure. We follow the exact same procedure as the main manuscript. We apply a chirp signal ranging between 0.3 kHz and 8 kHz to the thickest HEC and measure the velocity of each HEC. We see a high transmission area (the bright color) with a negative slope, as shown in Fig.~\ref{fig:sweep_thinning}(a). Unlike the positive gradient HEC chain (Fig.~\ref{fig:sweep}(a)), the transmission level is kept high to the end of the chain. The HEC mass decreases towards the chain end, resulting in higher particle velocity to meet the conservation of momentum. 
This agrees well with the FEA result in Fig.~\ref{fig:sweep_thinning}(b). 
The high transmission in Fig.~\ref{fig:sweep_thinning}(a) and (b) corresponds to the first pass band in the band structure in Fig.~\ref{fig:sweep_thinning}(c). This band structure is a mere flip in the horizontal direction of Fig.~\ref{fig:DispCurve}(c). Due to the thick stop band from the thick HECs, wave cannot propagate through the chain at high frequency range (4.5 kHz - 8 kHz). For the same reason, the wave in the low frequency range stops traveling once they hit this boundary. In other words, the wave is reflected against the first stop band and thus, is not transmitted through the chain.

\begin{figure}[ht]
\centering
 \includegraphics[width=1\linewidth]{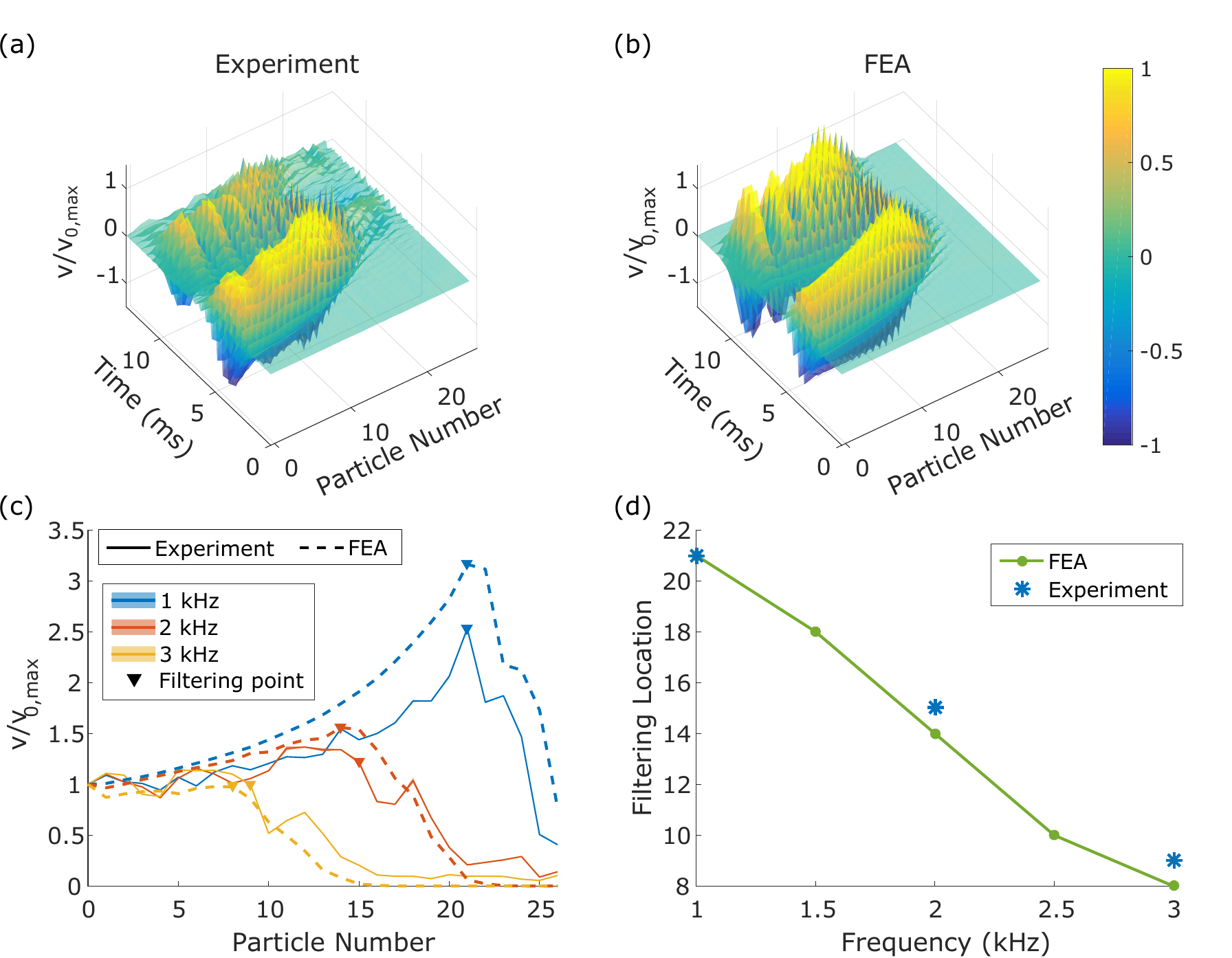}
 \caption{Velocity response at the input frequency of 2 kHz mapped in time and space domain, from (a) the experiment and (b) the FEA simulation of the negatively-graded HEC chain. The velocity is normalized by the maximum input velocity ($v_{0,max}$). (c) The maximum normalized velocity of the propagating wave (excluding reflected wave) at each HEC. The blue, orange, and yellow solid lines indicate the input frequency of 1, 2, and 3 kHz, respectively. The solid lines represent experimental result and the dashed lines represent FEA result. The shaded area, although hardly visible, represent the standard deviation from the experiment. We find the point where the wave amplitude starts to decline and mark with inverted triangles. These locations are plotted with respect to the corresponding input frequency in (d). The experimental results are plotted as asterisk markers and the FEA results are plotted in the green solid lines with circular markers.}
 \label{fig:Vmap_thinning}
\end{figure}

We verify the reflection effect of the negatively-graded HEC chain at different frequencies. We excite the chain with a Gaussian pulse at 2 kHz and measure the velocity response. As seen in Fig.~\ref{fig:Vmap_thinning}(a), the input wave quickly dies and reflects back at around 21st HEC. We observe a very similar behavior in FEA result in Fig.~\ref{fig:Vmap_thinning}(b). This is so-called boomerang or rainbow effect, which has been reported in \cite{Arreola-Lucas2017,Tsakmakidis2007}. 
The quantitative difference of amplitude between the experimental and computational results is due to the friction and damping effects. 

We measure the transient response at other input frequencies and plot the maximum velocity at each HEC in Fig.~\ref{fig:Vmap_thinning}(c). We notice that the reflection location comes close to the excitation point at the higher input frequency. In other words, the wave is reflected earlier at the higher excitation frequency. We also find that the maximum velocity amplitude increases towards the chain end. As mentioned in the previous paragraph, the decreasing mass plays a role to this amplified response. We confirm again that the experiment (solid lines) and the FEA (dashed lines) are in good agreement. 
We plot the reflection point with respect to the input frequency in Fig.~\ref{fig:Vmap_thinning}(d). We can clearly see the negative relationship between the reflection location and the input frequency.  Moreover, the filtering locations match the transmission boarder as indicated by the star and circular marks in Fig.~\ref{fig:sweep_thinning}(a) and (b), respectively. 


\clearpage
\bibliographystyle{unsrt}
\bibliography{reference}

\end{document}